\documentclass[onecolumn,showpacs,floatfix,aps,nofootinbib,11pt]{revtex4-1}
\usepackage{amssymb,amsmath}
\usepackage[dvips]{graphics,color}
\usepackage{epsfig}\usepackage{float}
\usepackage{bm,accents}
\newcommand{\mais}{{\textsc{\tiny +}}}
\newcommand{\menos}{{\textsc{\tiny $-$}}}
\RequirePackage{amssymb}
\RequirePackage{amsmath}
\usepackage{makeidx}
\def\dual#1{\accentset{\boldsymbol{\neg}\vspace{-.13ex}}{#1}}

\def\dual#1{\accentset{\boldsymbol{\neg}\vspace{-.13ex}}{#1}}

\newcommand{\beq}{\begin{eqnarray}}\newcommand{\benu}{\begin{enumerate}}\newcommand{\enu}{\end{enumerate}}
\newcommand{\eeq}{\end{eqnarray}}
\usepackage[brazil]{babel}
\usepackage[latin1]{inputenc}
\usepackage{graphicx}
\usepackage{indentfirst}
\usepackage{color}
\newcommand{\be}{\begin{equation}}
\newcommand{\ee}{\end{equation}}

\newcommand{\ba}{\begin{eqnarray}}
\newcommand{\ea}{\end{eqnarray}}

\DeclareMathOperator{\e}{e}
\def\dual#1{\accentset{\boldsymbol{\neg}\vspace{-.13ex}}{#1}}

\def\0{\bm0}
\def\vp{\bm\varphi}

\def\kb{\bm\kappa}

\def\s{\bm\sigma}
\def\p{\bm{p}}
\def\0{\bm0}
\def\vp{\bm\varphi}

\def\kb{\bm\kappa}

\def\s{\bm\sigma}
\def\p{\bm{p}}

\begin{document}

\title{ Dark Spinors  
Hawking Radiation  in  String Theory Black Holes}
\author{R. T. Cavalcanti}
\email{rogerio.cavalcanti@ufabc.edu.br}
 \affiliation{
CCNH, 
Universidade Federal do ABC, 09210-580, Santo Andr\'e - SP, Brazil}
 \affiliation{
Dipartimento di Fisica e Astronomia, Universit\`a di Bologna, 
via Irnerio 46, 40126 Bologna, Italy}
\author{Rold\~ao da Rocha}
\email{roldao.rocha@ufabc.edu.br} \affiliation{
CMCC, 
Universidade Federal do ABC, 09210-580, Santo Andr\'e - SP, Brazil.}
\affiliation{International School for Advanced Studies (SISSA), Via Bonomea 265, 34136 Trieste, Italy}
\pacs{04.70.Dy, 04.50.Gh, 11.25.Wx}

\begin{abstract}
\begin{center}
\textsc{Abstract}
\end{center}
The Hawking radiation spectrum of Kerr-Sen axion-dilaton black holes is derived, in the context of dark spinors  tunnelling across the horizon. Since a black hole has a well defined temperature,  it should radiate in principle all the standard model particles, similar to a 
black body at that temperature.  We investigate the tunnelling of mass dimension one spin-1/2  dark fermions, that are beyond the standard model and are prime candidates to the dark matter. Their  interactions with the standard model  matter and gauge fields are suppressed by at least one power of unification scale, being  restricted just to the Higgs field and to the graviton likewise. 
The tunnelling method 
for the emission and absorption of mass dimension one particles across the event horizon of Kerr-Sen axion-dilaton black holes is shown here to provide further evidence for the universality of black hole radiation, further encompassing particles beyond the standard model.  \end{abstract}
\maketitle

\section{Introduction}
Black hole tunnelling procedures have been placed as prominent   methods to calculate the  temperature  of black holes \cite{1,2,3,5,6,7,8,12,Vanzo:2011wq}.
Tunnelling methods provide models for describing   the black hole radiation. Various types of black holes have been investigated in the context of tunnelling of fermions and bosons as well \cite{1,Vanzo:2011wq,2,k1}.
Complementarily to the first results that studied  the  tunnelling of particles across black holes \cite{1,2}, the Hamilton-Jacobi method was employed  \cite{3} and  further generalized, by applying the WKB approximation to the Dirac equation \cite{k1}.
Tunnelling procedures are quite used to investigate   black holes radiation, by taking into account classically forbidden paths that particles go through,  from the inside to the outside of  black holes. Moreover, quantum WKB approaches were employed to calculate  corrections to the Bekenstein-Hawking entropy for the Schwarzschild black hole \cite{Banerjee:2009wb}. 

The tunnelling method was employed to provide Hawking radiation due to dark spinors for black strings \cite{meuepl}. Moreover, this method was also employed to model the emission of spin-1/2 fermions, {}{and the Hawking  radiation  was deeply analyzed as the tunnelling of Dirac particles throughout an event horizon, where quantum corrections in the single particle action are proportional to the usual semiclassical contribution. In addition, the modifications to the Hawking temperature and Bekenstein-Hawking entropy were derived for the Schwarzschild black hole.} When   spin-1/2 fermions are taken into account,  the effect of the spin of each type of fermion cancels out, due to particles with the spin in any direction. Hence, the lowest WKB order implies that the black hole intrinsic angular momentum  remains constant in tunnelling processes. 
Hawking radiation emulates semi-classical quantum tunnelling methods, wherein the Hamilton-Jacobi method is comprehensively used \cite{3}. Mass dimension 3/2 fermions tunnelling  have been studied in the charged dilatonic black hole, the rotating Einstein-Maxwell-Dilaton-Axion black hole and the rotating Kaluza-Klein black hole likewise \cite{Chen:2008vi}. For a review see Ref. \cite{Vanzo:2011wq}.

Our approach here is to employ an exact classical solution in the low energy effective field theory describing heterotic string theory: the Kerr-Sen axion-dilaton black holes \cite{sen}. They present charge, magnetic dipole moment, and angular momentum,    involving the antisymmetric tensor field coupled to the Chern-Simons 3-form.
 A myriad  of black holes has been considered for tunnelling methods of fermions and bosons, as rotating and accelerating black holes, topological, BTZ, Reissner-Nordstr\"om, Kerr-Newman, and Taub-NUT-AdS black holes, including also the tunnelling of higher spin fermions  as well \cite{yale}. 
 
 We shall study similar methods for spin-1/2 fermions of mass dimension one, different of the procedure for standard mass dimension 3/2 fermions. 
Elko dark spinors, namely the dual-helicity eigenspinors of the charge conjugation operator \cite{allu,crevice}, are  spin-1/2  fermions of mass dimension one, with  novel features that make them capable to  incorporate both the Very Special Relativity (VSR) paradigm and the dark matter description as well \cite{allu,al2}. Such spinors  seem to be indeed a tip of the iceberg for a comprehensive class of non-standard (singular) spinors \cite{daRocha:2013qhu,Cavalcanti:2014wia}.
Moreover,  a mass generation mechanism has been introduced in  \cite{alex} for such dark particles,  by a natural coupling to the {kink} solution in field theory. 
It provides exotic couplings among scalar field topological solutions and Elko dark spinors    \cite{alex,daSilva:2012wp}. 
 Due to its very small coupling with the standard model fields, except  the Higgs field, dark spinors 
supply natural self-interacting dark matter prime candidates. Except for scalar fields and gravity, Elko dark spinors interactions with the standard model matter and gauge fields is suppressed by at least one power of unification/Planck scale \cite{allu}. In fact, the Lagrangian of such a field contains a quartic self interaction term and the interaction term of the new field with spin-zero bosonic fields. Moreover,   Elko framework is shown to be invariant under the action of the HOM(2) VSR group and covariant under SIM(2) VSR group \cite{cohen}.    Elko dark spinor fields is a representative of mass dimension one spin-1/2 fermions in the type-5 spinor field class in Lounesto's spinors classification, however is not the most general, since Majorana spinors  are also encompassed by such class \cite{daSilva:2012wp}. Some attempts to detect Elko  at the LHC have been proposed, and   important applications to cosmology  have been widely investigated as well \cite{meuepl,allu,al2,boehmer,jhep,S.:2014dja}. 

This article is presented as follows: the Kerr-Sen axion-dilaton black hole is briefly revisited in the next section, together with the dark spinors  framework. We thus shall  
calculate in Section III the probabilities of emission and absorption 
of Elko dark particles  across these black holes. Therefore the WKB approximation is used to compute  the tunnelling rate and thus the resulting tunnelling probability. Finally the associated Hawking temperature 
shall be obtained, corroborating the universal character of the Hawking effect and further extending it to particles beyond the standard model.

\section{Kerr-Sen axion-dilaton black holes and dark spinors}

String theory has solutions describing extra-dimensional extended objects surrounded by event horizons, presenting a causal structure associated to  singularities in string theory. 
The low energy effective action of the heterotic
string theory is ruled by an action that, up to higher
derivative terms and other fields which are set to zero for the
particular class of backgrounds  considered \cite{sen}, is given by
\begin{eqnarray}
S&=&-\!\int d^4 x\sqrt{-\det G} e^{-\Phi} \left(-R+\frac{1}{12}H_{\mu\nu\rho}H^{\mu\nu\rho}-G^{\mu\nu}\partial_\mu\Phi\partial_\nu\Phi
+\frac18 F_{\mu\nu}F^{\mu\nu}\right)
\label{acao}
\end{eqnarray}
where  $G_{\mu\nu}$ is the metric regarding a $\sigma$-model \cite{sen}, related to the Einstein metric
by  $
e^{-\Phi} G_{\mu\nu}$, $\Phi$ denotes the dilaton field, $R$ stands for the scalar curvature,
$F_{\mu\nu}=\partial_{[\mu} A_{\nu]}$ is the Maxwell  field strength,  
and 
$H_{\mu\nu\rho}=\partial_\mu B_{\nu\rho} + \partial_\rho B_{\mu\nu} + \partial_\nu B_{\rho\mu} - 
\Omega_{\mu\nu\rho}$, for the
Chern-Simons 3-form $\Omega_{\mu\nu\rho}-\frac14 A_{(\mu} F_{\nu\rho)}$. The above action can be led to the one in \cite{1112}, up to the $H_{\mu\nu\rho}H^{\mu\nu\rho}$ term, after field redefinition.

The Kerr-Sen dilaton-axion black hole metric is a solution of the field equations derived from (\ref{acao}). In Boyer-Lindquist coordinates  it reads:
\begin{align}\nonumber
ds^{2} =&-\frac{\Delta-a^2\sin^2\theta}{\Sigma}dt^2+\frac{\Sigma}{\Delta}dr^2+\Sigma d\theta^2+\frac{\sin^2\theta}{\Sigma}\left[\left(r^2-2\beta r -a^2 \right)^2-\Delta a^2\sin^2\theta \right]d\varphi^2+\nonumber\\
&-\frac{2a\sin^2\theta}{\Sigma}\left[\left(r^2-2\beta r -a^2 \right)^2-\Delta \right]dtd\varphi\label{metricaaa}
\end{align}
where
\begin{align}
\!\!\!\!\Sigma=r^2-2\beta r +a^2\cos^2\theta,\qquad \Delta=r^2 - 2\eta r + a^2 = (r - r_\mais )(r - r_\menos )\qquad \beta = \eta \sin h^2 \frac{\alpha}{2}
\end{align} and 
$r_\mais [r_\menos]$ are the coordinate outer [inner] singularities.

The metric (\ref{metricaaa}) describes a black hole solution with charge $Q$, mass $M$,  magnetic dipole moment $\mu$, and 
angular momentum $J$, given by
\begin{eqnarray}
Q&=&{\eta \over\sqrt 2}\sinh\alpha,~~~M={\eta\over 2} (1+\cosh\alpha), ~~~\mu ={1\over\sqrt 2} \eta a\sinh\alpha\,,~~~ J=
{\eta a\over 2} (1+\cosh\alpha)\,.
\end{eqnarray}
The associated $g$-factor can be expressed as
$
g={2\mu M\over QJ} = 2
$  \cite{1112}.
The  parameters can be expressed in terms of genuinely physical quantities as
$$
\eta=M-{Q^2\over 2M}, ~~~~~~\alpha =\arcsin\!{\rm h}\left({2\sqrt 2 QM\over 2M^2 -Q^2}\right), ~~~~~~
a={J\over M}
$$
The coordinate singularities thus read
$
r_\pm = M-{Q^2\over 2M}\pm\sqrt{-{J^2\over M^2}+\left(M-{Q^2\over 2M}\right)^2
}
$ which vanishes unless $
|J| < M^2-\frac{Q^2}{2}
$. 
The area of the outer event horizon is given by $$
A=8\pi M \left(M-{{Q^2\over 2M}} +\sqrt{-{J^2\over M^2}+\left(M-{Q^2\over 2M}\right)^2
}\right)\,.
$$
Thus in the extremal limit, since 
$|J|\to M-{Q^2\over 2M}$, it reads  
$A\to 8\pi |J|$. In this limit the horizon is hence finite and the surface gravity $\kappa$, or equivalently, the Hawking temperature
$T_H=\kappa/2\pi$ is provided by \cite{sen}
$$
\kappa
={\sqrt{ -4J^2+(2M^2 -Q^2)^2}\over 2M(2M^2 -Q^2 +\sqrt{ -4J^2+(2M^2-Q^2)^2})}
$$
Thus in the extremal limit we have the limit $\kappa\to 0$ if $J\ne 0$.
On the other hand, if $J=0$ then $\kappa= {1\over 4M}$,
in agreement with the results of refs. \cite{1111,1112}.
For $J\ne 0$ this black hole solution has aspects analogous to the extremal rotating
black hole rather than extremal charged black holes  \cite{1112}. 

By performing the transformation  $\phi = \varphi -\Omega t$, where
$\Omega=\frac{a\left(a^2-2\beta r +a^2-\Delta\right)}{\left(r^2-2\beta r+a^2\right)^2-\Delta a^2\sin^2\theta}$, 
the metric (\ref{metricaaa}) takes the form
\begin{align}\label{metrrr}
\!\!\!\!\!ds^{2} =\!-\frac{\Delta\Sigma}{\left(r^2\!-\!2\beta r\!-\!a^2 \right)^2\!-\!\Delta a^2\sin^2\theta}dt^2\!+\!\frac{\Sigma}{\Delta}dr^2\!+\!\Sigma d\theta^2\!+\!\frac{\sin^2\theta}{\Sigma}\left[\left(r^2\!-\!2\beta r \!-\!a^2 \right)^2\!-\!\Delta a^2\sin^2\theta \right]d\phi^2\,.
\end{align}To study the Hawking radiation at the event horizon,  the metric is regarded near the horizon:
\begin{align}
ds^2=-F(r_\mais)dt^2+\frac{1}{G(r_\mais)}dr^2+\Sigma(r_\mais)d\theta^2+\frac{H(r_\mais)}{\Sigma(r_\mais)}d\phi^2
\end{align}
where
\begin{align}
H(r_\mais)&= \sin^2\theta\left(r_\mais^2-2\beta r_\mais +a^2 \right)^2,\qquad 
F(r_\mais)= \frac{2(r_\mais-\eta)(r-r_\mais)\Sigma(r_\mais)}{\left(r_\mais^2-2\beta r_\mais+a^2\right)^2}\\
G(r_\mais)&= \frac{2(r_\mais-\eta)(r-r_\mais)}{\Sigma(r_\mais)}\,.
\end{align}
In order to analyze the tunnelling of Elko dark particles  across the Kerr-Sen black hole event horizon, we will study the role that Elko dark particles  play in this background. The essential prominent Elko
 particles features  are in short 
revisited \cite{allu}. Elko dark spinors  $\lambda(p^\mu)$ are eigenspinors of the charge
conjugation operator $C$, namely, $C\lambda(p^\mu)=\pm \lambda(p^\mu)$. 
The plus [minus] sign regards {self-conjugate},  [{anti self-conjugate}]  spinors , denoted by $\lambda^{S}(p^\mu)$ [$\lambda^{A}(p^\mu)$]. For rest spinors  $\lambda(k^\mu)$ the boosted spinors read $
	\lambda(p^\mu) = e^{i \kb\cdot\vp} \lambda(k^\mu),$ where $
		k^\mu = (m,\lim_{p\rightarrow 0}{\p}/{\vert\p\vert}),$ where  $e^{i \kb\cdot\vp}$ denotes the boost operator. 
The $\phi_{}(k^\mu)$ are defined to be  eigenspinors of  the helicity operator, as $
\s\cdot\hat \p\, \phi_{}^\pm(k^\mu) = \pm \phi_{}^\pm(k^\mu)$, 
where \cite{allu} 
\begin{eqnarray}
\phi^+_{}(k^\mu) &=& \sqrt{m} \left(
									\begin{array}{c}
									\cos\left(\frac{\theta}{2}\right)e^{- i \varphi/2}\\
									\sin\left(\frac{\theta}{2}\right)e^{+i \varphi/2}
											\end{array}\right)\equiv\left(\begin{array}{c}
\alpha\\\beta			\end{array}
									\right)\,, \label{phim}\\ 
\phi^-_{}(k^\mu) &=& \sqrt{m} \left(
									\begin{array}{c}
									-\sin\left(\frac{\theta}{2}\right)e^{- i \varphi/2}\\
									\cos\left(\frac{\theta}{2}\right)e^{+i \varphi/2}
											\end{array}
									\right)=\left(\begin{array}{c}
-\beta^*\\\alpha^*			\end{array}
									\right)\,. 	\label{phime}					\end{eqnarray}
Elko dark spinors   $\lambda(k^\mu)$ are constructed   as 
\begin{eqnarray}
 \lambda^S_\pm(k^\mu) & =& \left(
					\begin{array}{c}
					\sigma_2\left(\phi_{}^\pm(k^\mu)\right)^\ast\\
								\phi_{}^\pm(k^\mu)
					\end{array}
					\right)\,,\qquad\qquad  
				\lambda^A_\pm(k^\mu) = \pm\left(
							\begin{array}{c}
								-\sigma_2\left(\phi_{}^\mp(k^\mu)\right)^\ast\\
								\phi_{}^\mp(k^\mu)
												\end{array}
												\right)\,,
			\label{ppm}
\end{eqnarray}
and have dual helicity, as $ -i\sigma_2(\phi^\pm)^\ast$ has helicity dual to that of 
$\phi^\pm$. The boosted terms  \begin{eqnarray}
		\lambda^A_\pm(p^\mu) = \sqrt{\frac{E+m}{2 m} }\left( 1\pm\frac{p^\mu}{E+m}\right)\lambda^A_\pm\,,\qquad	\lambda^S_\pm(p^\mu) = \sqrt{\frac{E+m}{2 m} }\left( 1\mp\frac{p^\mu}{E+m}\right)\lambda^S_\pm 		\label{jj}
		\label{jj1}
\end{eqnarray}
 are the expansion coefficients of  a mass dimension one quantum field. The  Dirac operator does not annihilate the $\lambda(p^\mu)$, but instead the equations of motion read  \cite{allu,crevice}:
\begin{align}\label{elkodyn}
\gamma_\mu \nabla^\mu \lambda^S_\pm&=\pm i\frac{m}{\hslash} \lambda^S_\mp\\
\gamma_\mu \nabla^\mu \lambda^A_\mp&=\pm i\frac{m}{\hslash} \lambda^A_\pm\label{4}
\end{align}
Dark spinors  nevertheless satisfy the Klein-Gordon equation.

A mass dimension one quantum field can be thus constructed as \cite{crevice} 
\begin{eqnarray}
\!\!\!\!\!\!\mathfrak{f}(x) =  \int \frac{\text{d}^3p}{(2\pi)^3}  \frac{1}{\sqrt{2 m E(\p)}} \sum_\rho \Big[ b^\dagger_\rho(\p)\lambda^A(\p) e^{i p_\mu x^\mu}+a_\rho(\p)\lambda^S(\p) e^{- i p_\mu x^\mu}{\Big]}\,.
\end{eqnarray}
The creation and annihilation operators $a_\rho(\p), a^\dagger_\rho(\p)$ satisfy the  Fermi statistics  \cite{crevice}, 
with similar anticommutators for $b_\rho(\p)$ and 
$b^\dagger_\rho(\p)$.
The mass dimensionality of $\mathfrak{f}(x)$ can be realized from the adjoint
 \begin{eqnarray}
\!\!\!\!\!\dual{\mathfrak{f}}(x)  \!=\!  \int \frac{{d}^3{\bf p}}{(2\pi)^3}  \frac{1}{\sqrt{2 m E(\p)}} \sum_\rho \Big[b_\rho(\p)\dual{\lambda}^A(\p) e^{-i p_\mu x^\mu} +  a^\dagger_\rho(\p)\dual{\lambda}^S(\p) e^{i p_\mu x^\mu} {\Big]}
\end{eqnarray}
where, by denoting hereupon  $\sigma^\mu$ the Pauli matrices,  
 $\dual{\lambda}_\rho(p^\mu)   = {\big[}\Xi\, \lambda_\rho(p^\mu){\big]}^\dagger \,\sigma_1\otimes\mathbb{I}_2,$ and 
$
\Xi = \frac{1}{2 m} \Big(\lambda^S_-(p^\mu)\bar\lambda^S_-(p^\mu)  -\lambda^A_-(p^\mu)\bar\lambda^A_-(p^\mu)+\lambda^S_+(p^\mu)\bar\lambda^S_+(p^\mu) - \lambda^A_+(p^\mu)\bar\lambda^A_+(p^\mu)\Big)\,$   is an involution \cite{crevice}, 
for the Dirac dual  $\bar\lambda(p^\mu){=} \lambda(p^\mu)^\dagger \gamma^0$. 
The mass dimension of the new field is determined by the SIM(2) covariant propagator \cite{crevice}
\begin{eqnarray}
\!\!\!\!S(x-x^\prime)  = i \left\langle\hspace{4pt}\left\vert {{\rm T}} \left( \mathfrak{f}(x) \dual{f}(x^\prime)\right)\right\vert\hspace{4pt}\right\rangle=- \lim_{\epsilon \to 0^+}\int\frac{{d}^4 p}{(2 \pi)^4} 
\e^{- i p^\mu \left(x^{\prime}_\mu - x_\mu\right)} \left(  \frac{ \mathbb{I} + \mathcal{G}(\varphi)}{p^\mu p_\mu - m^2 + i \epsilon} \right)
\end{eqnarray}
where ${\rm T}$ is the canonical time-ordering operator and  
\begin{equation}
\mathcal{G}(\varphi)  {=}
\left(\begin{array}{cc}
			 0 & -i e^{-i\varphi} \\
			i  e^{i\varphi} & 0 
			\end{array}\right)\otimes\sigma_1\,
\end{equation} that respects symmetries of
the theory of VSR \cite{cohen,crevice}.

\section{Hawking radiation from tunnelling dark spinors}

Hawking radiation from general black holes  encompasses distinguished 
charged and uncharged particles. Tunnelling methods can be employed for Elko dark particles  across  the horizon of Kerr-Sen black holes. From (\ref{metrrr})  the associated tetrad can be chosen in order to 
 the following generators  can be achieved:
\begin{align}
\gamma^t&=
\frac{1}{\sqrt{F(r_\mais)}}\left(\begin{array}{cc}
0 & \mathbb{I}_2 \\ 
\mathbb{I}_2 & 0
\end{array}\right)\,,\quad\quad   \gamma^\theta=
\frac{1}{\sqrt{\Sigma(r_\mais)}}\left(\begin{array}{cc}
0 & \sigma^2 \\ 
-\sigma^2 & 0
\end{array}\right)\,, \\
\gamma^r&
={\sqrt{G(r_\mais)}}\left(\begin{array}{cc}
0 & \sigma^1 \\ 
-\sigma^1 & 0
\end{array}\right)\,,\quad\quad \gamma^z=
\sqrt{\frac{\Sigma(r_\mais)}{H(r_\mais)}}\left(\begin{array}{cc}
0 & \sigma^3 \\ 
-\sigma^3 & 0
\end{array}\right)\,.
\end{align}

Elko dark spinors can be written as 
\begin{eqnarray}\label{elkof1}
\lambda^\mathrm{S}_+ &=&\!\left(
\begin{array}{c}
- i\beta^*  \\
i\alpha^* \\
\alpha \\
\beta
\end{array}%
\right) \exp \left( \frac{{}i{} }{\hbar }\tilde{I}\right)\,,\qquad 
\label{elkof2}\lambda^\mathrm{S}_- =\left(
\begin{array}{c}
- i\alpha  \\
-i\beta \\
-\beta^* \\
\alpha^*
\end{array}%
\right) \exp \left( \frac{{}i{} }{\hbar }\tilde{I}\right)\,,\\
\label{elkof3}\lambda^\mathrm{A}_+   &=&\left(
\begin{array}{c}
i\alpha  \\
i\beta \\
-\beta^* \\
\alpha^*
\end{array}%
\right) \exp \left( \frac{{}i{} }{\hbar }\tilde{I}\right)\,,\qquad\lambda^\mathrm{A}_-  =\left(
\begin{array}{c}
- i\beta^*  \\
i\alpha^* \\
-\alpha \\
-\beta
\end{array}%
\right) \exp \left( \frac{{}i{} }{\hbar }\tilde{I}\right)\,,
\end{eqnarray}
where   $\tilde{I}=\tilde{I}(t,r,\theta,z)$ represents the classical action.  We use the above forms for  dark particles    in each one of the Eqs. (\ref{elkodyn}-\ref{4}), and then solve this coupled system of  equations. 
By denoting  
$$
\nabla_{\mu }=\partial _{\mu }+\frac{1}{8}{}i{} \Gamma _{\mu }^{\alpha \beta }\left[ \gamma
^{\alpha },\gamma ^{\beta }\right],$$ where  $\gamma ^{\sigma}$ are the usual Clifford bundle generators for the Minkowski spacetime. By identifying  $\lambda$ [$\mathring{\lambda}$] to the Elko spinor on the left [right] hand side of Eqs. \eqref{elkodyn} and \eqref{4}, then Eq. \eqref{elkodyn} reads $\gamma^\mu (\nabla_\mu +eA_\mu)\lambda=i\frac{m}{\hslash} \mathring{\lambda}$. Using the WKB approximation, where $\tilde{I}=I+\mathcal{O}(\hslash)$, it yields    
\begin{align}\label{elkoelko}
({{I}_\mu}+eA_\mu)\gamma^\mu \lambda=i{m} \mathring{\lambda}+\mathcal{O}(\hslash)\,,
\end{align}
where ${I}_\mu\equiv\frac{\partial {I}}{\partial x^\mu}$. Taking merely the leading order terms in the above equation, from a general form 
$\lambda=(a,b,c,d)^\intercal$, $  \mathring{\lambda}=(\mathring{a}, 
\mathring{b}, \mathring{c}, 
\mathring{d})^\intercal\,,$
we have general Elko dynamic equations governed by (\ref{elkoelko})
The ansatz $I(t,r,\theta,\varphi)=-(\omega-j\Omega)t+j\varphi+W(r)+\Theta(\theta)$ can be  used, where $\omega$ and $j$ denote the energy and magnetic quantum number of the particle respectively. 
Moreover, the parameters $a,b,c,d$  are not independent. In fact, Eqs. (\ref{elkof1}) and   (\ref{elkof3}) assert that for the self-conjugate spinors  $\lambda^S$ we have $a=-id^*$ and $b=ic^*$, whereas for the anti-self-conjugate spinors $\lambda^A$ it reads $a=id^*$ and $b=-ic^*$. Thus, by corresponding the $\lambda^S$  [$\lambda^A$] spinors to the upper [lower] sign below, after awkward computation Eq.(\ref{elkoelko}) yields
\begin{align}\label{s1}
\pm \sqrt{G(r_\mais)}W' d^* \mp \frac{(\omega-j\Omega_H+eA_+)}{\sqrt{F(r_\mais)}} c^* =\,\mathring{d}\,m\\ 
\label{s4}
\pm  \sqrt{G(r_\mais)}W' c^*\mp \frac{(\omega-j\Omega_H+eA_+)}{\sqrt{F(r_\mais)}} d^*=-\,\mathring{c}\,m\,.
\end{align}
The angular function $j\varphi +\Theta(\theta)$ must be a complex function and the same solution for it is  achieved for both the incoming and outgoing cases as well. It implies that the contribution of such function vanishes after dividing the outgoing probability by the incoming one. Hence the angular function can be neglected hereupon.

The above system the equations for $\lambda^S_+$ [$\lambda^S_-$] are shown to be  equivalent to the ones for $\lambda^A_-$ $[\lambda^A_+$]. Thus we have to deal solely with the self-conjugate $\lambda^S_\pm$ spinors. Moreover, there are more underlying equivalences. In fact,  Eq. \eqref{s1} for $\lambda^S_\pm$ is equivalent to Eq. \eqref{s4} for $\lambda^S_\mp$. Consequently there is just a  couple of equations  for $\lambda^S_\pm$ 
given by:
\begin{align}\label{ss1}
\left\{\begin{array}{ccc}
\sqrt{G(r_\mais)} W'\alpha^*-\frac{(\omega-j\Omega_H+eA_+)}{\sqrt{F(r_\mais)}}\beta^* & = & \beta^* m \\ 
\sqrt{G(r_\mais)} W'\beta+\frac{(\omega-j\Omega_H+eA_+)}{\sqrt{F(r_\mais)}}\alpha & = & \alpha m
\end{array} \right.\\\label{ss2}
 \left\{\begin{array}{ccc}
\sqrt{G(r_\mais)} W'\alpha\mp\frac{(\omega-j\Omega_H+eA_+)}{\sqrt{F(r_\mais)}}\beta & = & \mp\beta m \\ 
\sqrt{G(r_\mais)} W'\beta^*\mp\frac{(\omega-j\Omega_H+eA_+)}{\sqrt{F(r_\mais)}}\alpha^* & = &\pm \alpha^* m
\end{array} \right.\end{align}

Combining either Eqs. \eqref{ss1} or \eqref{ss2} implies equations   for either $\lambda^S_+$ or $\lambda^S_-$, respectively. Hence, for each $\lambda^S$ there is a system of  coupled equations for the dark spinor components $\alpha$ and  $\beta$ and also another coupled system for $\alpha^*$ and  $\beta^*$, which are going to be solved separately. We denote now the first equation of each one of the systems below to be the equations related to $(\alpha, \beta)$, whereas the second ones regard $(\alpha^*, \beta^*)$. We can determine the above functions as:
\begin{align}\label{sss1}
\lambda^S_+:& \left\{\begin{array}{cccc}
&W_1(r) & = & \pm \int\sqrt{\frac{m^2F-(\omega-j\Omega_H+eA_+)^2}{FG}}dr \\ 
&W_2(r) & = & \pm \int\sqrt{\frac{(m\sqrt{F}+\omega-j\Omega_H+eA_+)^2}{FG}}dr
\end{array} \right.\\\label{sss2}
\lambda^S_-:& \left\{\begin{array}{ccccc}
&W_3(r) & = & \pm \int\sqrt{\frac{(m\sqrt{F}-\omega+j\Omega_H+eA_+)^2}{FG}}dr \\ 
&W_4(r) & = & iW_2(r)
\end{array} 
\right.
\end{align}
For massless particles the solutions for $\lambda^S_\pm$ are equivalent, being given by
\begin{align}
\!\!\!\!\!W_2(r)=\pm\int\sqrt{\frac{(\omega-j\Omega_H+eA_+)^2}{FG}}dr=-iW_1(r)\,.
\end{align}
The above expression for $W_2(r)$ is similar to the results  in \cite{DE} when $m=0$. The solution for $W_2(r)$ is provided by
\begin{align}
W_2(r)=\pm i\pi \frac{r_\mais^2-2\beta r_\mais+a^2}{2(r_\mais+m)}(\omega-J\Omega_H+eA_+)\,.
\end{align}
Since $F\to 0$ near the black hole horizon, the results hold both 
for massive and their massless limit particles as well. 
It is worth to emphasize that it implies a real solution for $W_1(r)$, and thus a null contribution for the tunnelling effect. 

In compliance with the WKB approximation, the tunnelling rate reads $\Gamma \propto \exp(-2\,\mathtt{Im}\,I)$, where $I$ denotes the classical action for the path. Hence the imaginary part of the action becomes a prominent goal for the  tunnelling process. The imaginary part of the action  yields
\begin{align}
\mathtt{Im}\,I_{\pm}=\pm i\pi \frac{r_\mais^2-2\beta r_\mais+a^2}{2(r_\mais+m)}(\omega-J\Omega_H+eA_+)\,.
\end{align}
Thus the resulting tunnelling probability is given by
\begin{align}
\Gamma&=\frac{P_{({\rm emission})}}{P_{({\rm absorption})}}=\frac{e^{-2\mathtt{Im}I_+}}{e^{-2\mathtt{Im}I_-}}=-2\pi \frac{r_\mais^2-2\beta r_\mais+a^2}{(r_\mais+m)}(\omega-J\Omega_H+eA_+)\,.
\end{align}
Finally  the Hawking temperature of the Kerr-Sen dilaton-axion black hole is acquired:
\begin{align}\label{th}
T_H&=\frac{1}{2\pi} \frac{(r_\mais+m)}{r_\mais^2-2\beta r_\mais+a^2}
\end{align}
which is an universal formula also obtained by other methods for the Hawking temperature from fermions tunnelling \cite{DE}. Obviously when the parameter $a$ tends to zero,  Eq.(\ref{th}) provides the well-known Hawking radiation associated to the static black hole. 

\section{Concluding Remarks}
We derived the Hawking radiation for spin-1/2 fermions of mass dimension one, represented by  dark spinors  tunnelling across a  Kerr-Sen dilaton-axion black hole horizon. The temperature of these solutions were computed and demonstrated to  confirm the universal character of the Hawking effect, even for mass dimension one fermions of spin-1/2 that are beyond the standard model. The mass dimension one feature  of such spinor fields sharply suppresses the  couplings to other fields of the standard model. Indeed, by power counting arguments, Elko spinor fields can  perturbatively and renormalizably self-interact and further to a scalar (Higgs) field. This type of interaction means an unsuppressed quartic self interaction. The quartic self interaction is essential to dark matter  
observations \cite{10,101}. Therefore Elko spinor fields perform an adequate fermionic dark matter candidate.
It can represent a real model for dark matter tunnelling across black holes. Corrections of higher order in $\hbar$ to the Hawking temperature (\ref{th}) of type $I = I_0 + \sum_{n\geq 1} \hbar^n I_n$  \cite{Banerjee:2009wb} can be still implemented in the context of mass dimension one spin-1/2 fermions. Moreover, other mass dimension one fermions \cite{Cavalcanti:2014wia} and higher spin mass dimension one fermions can be studied in the context of black hole tunnelling methods,   however these issues are beyond the scope of this paper, that comprised dark particles   
tunnelling across  Kerr-Sen dilaton-axion black holes.

\acknowledgments
RTC thanks to CAPES and to UFABC.
RdR is grateful to CNPq grants No. 303027/2012-6, No. 451682/2015-7 and No. 473326/2013-2, for partial financial support, and to FAPESP Grant No. 2015/10270-0.

\end{document}